# Effect of Samarium doping on the nucleation of fcc-Aluminum in undercooled liquids

Yang Sun[1*], Feng Zhang[1], Lin Yang[1], Zhuo Ye[1], Huajing Song[1], Mikhail I. Mendelev[1], Cai-Zhuang Wang[1,2], Kai-Ming Ho[1,2,3†]

[1]Ames Laboratory, US Department of Energy, Ames, Iowa 50011, USA

[2]Department of Physics, Iowa State University, Ames, Iowa 50011, USA

[3]Hefei National Laboratory for Physical Sciences at the Microscale and Department of Physics, University of Science and Technology of China, Hefei, Anhui 230026, China

(Dated: May 2, 2018)

**Abstract**

The effect of Sm doping on the fcc-Al nucleation was investigated in Al-Sm liquids with low Sm concentrations ($x_{Sm}$) with molecular dynamics simulations. The nucleation in the moderately undercooled liquid is achieved by the recently developed persistent-embryo method. Systematically computing the nucleation rate with different $x_{Sm}$ ($x_{Sm}$=0%, 1%, 2%, 3%, 5%) at 700 K, we found Sm dopant reduces the nucleation rate by up to 25 orders of magnitudes with only 5% doping concentration. This effect is mostly associated with the increase in the free energy barrier with minor contribution from suppression of the attachment to the nucleus caused by Sm doping.

[*]E-mail: yangsun@ameslab.gov
[†]E-mail: kmh@ameslab.gov

Al-based alloys are widely used as materials for engineering and manufacturing due to their low density and high specific strength [1]. In addition, there is extensive interest [2–14] in glass forming alloys because of their superior mechanical properties. Among the class of glass-forming binary alloys, Al-rare-earth (Al-RE) amorphous alloys form a special subclass [15] because the glass formation range in Al-RE binary alloys lies on the solute-rich side of the eutectic point which violates the general rule that the binary metallic glass forms near the eutectic region [13,16,17]. In this paper, we focus on the Al-Sm system [11], a typical example of the glass forming Al-RE alloys[16], to study the effects of Sm doping on the nucleation of the face-centered cubic (fcc) Al crystals. Our study aims to elucidate the mechanism by which minor doping of RE elements can dramatically change the phase selection process by avoiding nucleation of fcc Al in the undercooled liquid[18].

Molecular dynamics (MD) simulation is an excellent tool that provides detailed information for phase transformations on the atomic level [19]. Recent MD simulations have revealed the nucleation kinetics in AlSm metallic glass [20] and the detailed growth kinetics of the devitrified AlSm crystal phases [21,22]. It was found that in the AlSm metallic glass, the Sm solute can retard Al nucleation by increasing the kinetic barrier to reduce the nucleus attachment rate [20]. Although the kinetic barrier can be the dominant factor controlling the nucleation in the glass state [20], it remains unclear how the Sm solutes affect the Al nucleation in a moderately undercooled liquid. On the other hand, due to the limitation of the simulation time, conventional MD simulations cannot access the nucleation event in the timescale which corresponds to the most common experimental conditions to measure the nucleation rate [23,24]. Here, we employ the recently developed persistent-embryo method (PEM) [25] to allow efficient sampling of rare Al nucleation events in the undercooled liquid. With the PEM, we are able to observe the Al nucleation without any biasing of the system when the critical nucleus forms which allows us to obtain accurate quantitative estimates of the critical nucleus size as a function of Sm doping. Within Classical Nucleation Theory (CNT), the simulation results can be used to compute the nucleation rate as a function of the Sm doping concentrations. Our results yield accurate quantitative estimates of contributions from thermodynamics versus kinetics effects of Sm addition.

According to the CNT [26], a homogeneous nucleation involves a formation of the critical nucleus in the undercooled liquid. The formation of such a nucleus is governed by two factors: one is the thermodynamic driving force towards the lower-free-energy bulk crystal. This term is negative and proportional to the nucleus size. The other is the energy penalty for creating an interface between the nucleus and the liquid. This term is positive and proportional to the area of the interface. Therefore, the excess free energy to form a nucleus with $N$ atoms is

$$\Delta G = N\Delta\mu + A\gamma \qquad (1),$$

where $\Delta\mu$ ($< 0$) is the chemical potential difference between the bulk solid and liquid, $\gamma$ is the solid-liquid interfacial free energy, and $A$ is the interface area which can be evaluated as $A = sN^{2/3}$ with $s$ being a shape factor. The competition between the bulk and interface terms leads to a nucleation barrier $\Delta G^*$ when the nucleus reaches the critical size $N^*$. CNT assumes the spherical shape for the nucleus to relate $\Delta G^*$ with $\gamma$ and $\Delta\mu$. We lift this assumption by introducing the shape factor $s$, assuming that the shape of the sub-critical nucleus does not change during growth. Mathematically, the interfacial free energy density $\gamma$ and the shape factor $s$, which are both difficult to compute, can be replaced by the critical nucleus size $N^*$ at the critical point[25] in the expression of the free energy barrier $\Delta G^*$, resulting in

$$\Delta G^* = \frac{1}{2}|\Delta\mu|N^* \qquad (2).$$

The crystal nucleation rate is the product of the probability to form the critical nucleus given by $\exp(-\Delta G^*/k_B T)$ and the kinetic prefactor. Following Auer and Frenkel[27], the expression of the nucleation rate can be written as:

$$J = \rho_L f^+ \sqrt{\frac{|\Delta\mu|}{6\pi k_B T N^*}} \exp(-\frac{|\Delta\mu|N^*}{2k_B T}) \quad (3)$$

where $f^+$ is the rate of single atom attachment to the critical nucleus and $\rho_L$ is the liquid density. Four factors, $\rho_L$, $N^*$, $\Delta\mu$, and $f^+$, are needed to compute the nucleation rate at a given temperature $T$. The liquid density $\rho_L$ can be obtained from the *NPT* simulation. The critical size $N^*$ and the attachment rate $f^+$ can be obtained from molecular dynamical simulations using the persistent-embryo method. The chemical potential difference $\Delta\mu$ is calculated separately using thermodynamic integration based on an alchemical path linking the pure Al liquid to the Al-Sm liquid [28].

A semi-empirical potential describing the interatomic interaction in the Al-Sm system was taken from Ref. [29]. This FS potential is developed to accurately reproduce the Al-Sm properties including the pure Al melting temperature and formation energies of the Al-Sm crystal phases. Previous studies have shown that this potential well describes the AlSm liquid structure similar to the *ab inito* MD simulations [18,30] and produces structure factors in good agreement with experimental measurements [31]. All MD simulations reported in the present paper were performed at T=700 K using the *NPT* ensemble with Nose-Hoover thermostat. The time step of the simulation was 1.0 *fs*. The simulation cell contained 13,500 atoms and was at least 20 times larger than the critical nucleus size. The initial liquid was equilibrated for 1 *ns*. All the simulations were performed using the GPU-accelerated LAMMPS code[32–34].

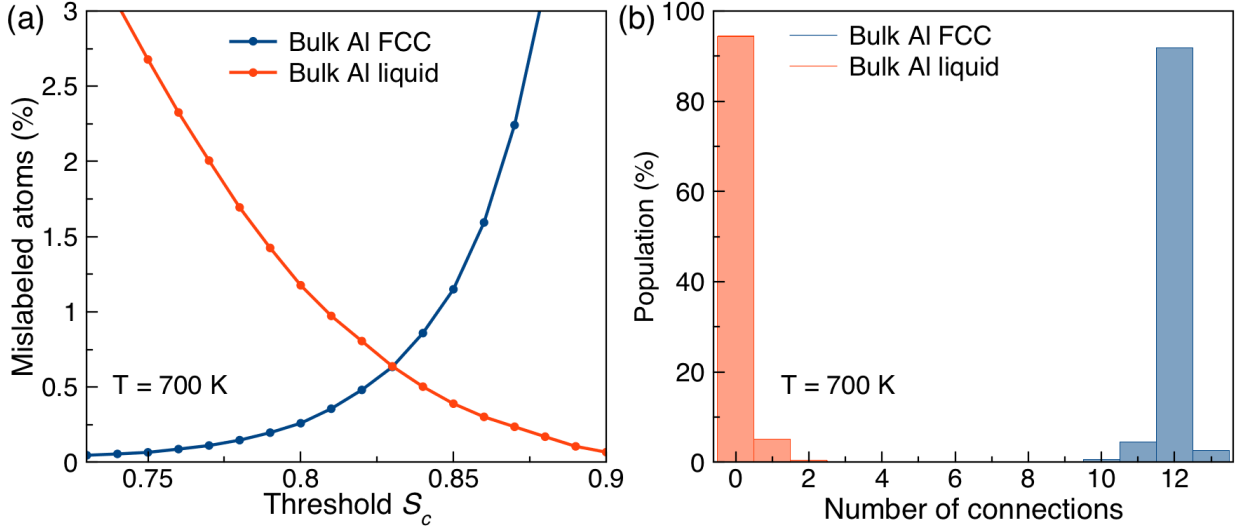

Fig. 1 Determination of the threshold to distinguish solid-like and liquid-like atoms. (a) Population of mislabeled atoms by different threshold values in bulk Al crystal and liquid at 700K. (b) Population of connections number per atom in bulk Al crystal and liquid at 700K.

To identify solid-like and liquid-like atoms during the MD simulation, the widely-used bond-orientational order (BOO) parameter [35,36] was employed by calculating $S_{ij} = \sum_{m=-6}^{6} q_{6m}(i) \cdot q_{6m}^{*}(j)$ between two neighboring atoms based on the Steinhardt parameter $q_{6m}(i) = \frac{1}{N_b(i)} \sum_{j=1}^{N_b(i)} Y_{lm}(\vec{r}_{ij})$, where $Y_{lm}(\vec{r}_{ij})$ is the spherical harmonics and $N_b(i)$ is the number of nearest neighbors of atom $i$. Two neighboring atoms $i$ and $j$ were considered to be connected when $S_{ij}$ exceeds a threshold $S_c$. To choose a statistically sound threshold value, we plotted the population of mislabeled atoms in bulk Al liquid and crystal as a function of the threshold values in Fig. 1(a). The crossing point between the mislabeling curves of the liquid and solid phases was chosen as the threshold [25,37]. With this threshold $S_c$, crystal and liquid can be well separated by counting the number of connections an atom has with its neighbors as shown in Fig. 1(b). We then used 6 as the connection cut-off to recognize solid-like atoms during the simulation [27]. The clustering analysis [38], which uses the crystalline bond length as the cutoff distance to choose neighbor atoms, is applied to measure the size of the nucleus formed during MD simulations. We examined the short-range order in the nucleus based on the packing motif of the atom cluster defined by the center atom and its nearest neighbor atoms. A cluster-alignment method [39] in which minimal root-mean-square deviations (RMSD) between the atom cluster and the perfect packing templates such as fcc, hcp and bcc polyhedra are calculated for crystal-structure recognition. The RMSD-based order parameter has been shown to be robust in characterization of the local structural ordering in the crystal, liquid and glasses [30,40,41].

In conventional MD simulation, the crystal nucleation is too rare event within the limited simulation time scales making it difficult to accurately measure the critical nucleus size except under extremely highly driven conditions[42]. The PEM allows efficient sampling of the nucleation process by preventing a small crystal embryo (with $N_0$ atoms which is much smaller than the critical nucleus) from melting using external spring forces[25]. This removes long periods of ineffective simulation where the system is very far away from forming a critical nucleus. As the embryo grows, the harmonic potential is gradually weakened and is completely removed when the cluster size reaches a sub-critical threshold $N_{sc}$ $(< N^*)$. During the simulation, the harmonic potential only applies to the original $N_0 (< N_{sc})$ embryo atoms. The spring constant of the harmonic potential can be expressed as $k(N) = k_0 \frac{N_{sc}-N}{N_{sc}}$ if $N < N_{sc}$ and $k(N) = 0$, otherwise. This strategy ensures the system is unbiased at the critical point such that a reliable value of $N^*$ is obtained. If the nucleus melts below $N_{sc}$ $(< N^*)$ the harmonic potential is gradually enforced preventing the complete melting of the embryo. When the nucleus reaches the critical size, it has equal chance to melt or to further grow causing critical fluctuations about $N^*$. Because the thermodynamic driving forces for growth or shrinking of the nucleus are smallest at the critical point, the $N(t)$ curve tends to display a plateau during the critical fluctuations giving us a unique signal to measure $N^*$. Moreover, in most cases, the critical nucleus forms several times before it successfully grows, giving us ample statistics to accurately determine $N^*$.

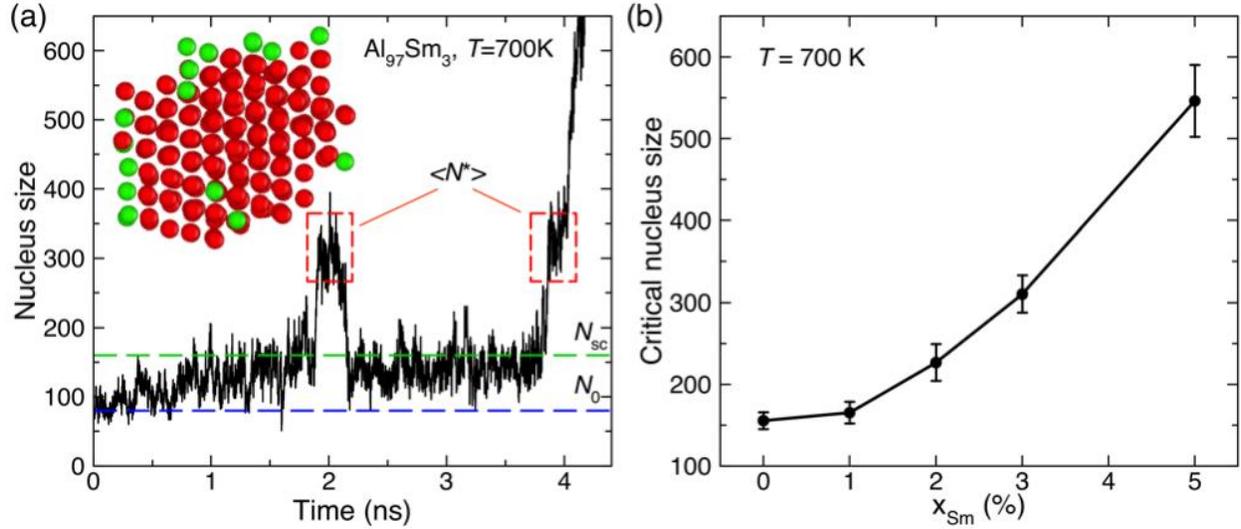

Fig. 2 The measurement of the critical nucleus size with PEM. (a) FCC nucleus size as a function of time in Al$_{97}$Sm$_3$ liquid at 700 K. The blue dashed line shows the atom number $N_0$ in the persistent embryo. The green dash line indicates the threshold $N_{SC}$ to remove the spring and the solid line indicates the critical size $N^*$. Two boxes (red) indicate the plateaus of critical nuclei. The insert shows a snapshot of the critical nucleus with red indicating FCC-like atoms and green indicating HCP-like atoms. (b) The critical nucleus size as a function of the Sm doping concentration in Al liquid at 700 K. The data points are connected to guide the eye. The error bar is obtained by calculating the standard deviation of the critical size obtained from multiple MD runs.

Figure 2(a) shows a trajectory simulated with PEM by using a pure-Al embryo in 3% Sm-doped Al liquid. The nucleus size fluctuates at the critical size region twice before it crosses the nucleation barrier to grow. Although the length of the fluctuating plateaus varies, the heights are almost identical. Thus, the critical size can be determined by averaging over the plateau heights. To obtain a statistically sound estimation of the critical size, multiple independent MD runs (up to 50) are performed to collect such critical plateaus. In Fig. 2(b), the obtained critical size is shown as a function of Sm concentration. With 1% Sm doping, the critical nucleus size is close to the one in the pure Al. However, when the doping concentration increases to 5%, the critical nucleus size becomes ~4 times larger. Analyzing the short-range order with the cluster-alignment method, the critical nucleus is fcc in structure with minor hcp-like atoms at the solid-liquid interface which is shown in Fig. 2(a).

Once the critical nucleus is obtained, we are able to compute the attachment rate at the critical nucleus. According to Auer and Frenkel [27,43,44], the attachment rate $f^+$ can be measured in MD simulations as the diffusion coefficient for the size change at the critical point by

$$f^+_{MD} = \frac{\langle |N(t)-N^*|^2 \rangle}{2t}. \quad (4)$$

As an example, we show in Fig. 3(a) the measurement of $f^+_{MD}$ in Al-3 at.%Sm liquid with the isoconfigurational ensemble [45] . 50 trajectories were generated from independent MD runs starting from the same atomic configuration containing a critical nucleus obtained by PEM, but with atomic velocities randomly assigned using the Maxwell distribution. Since the critical nucleus is at the top of the nucleation barrier, the nucleus eventually grows in half of the MD runs while melts in the other half runs, resulting in spreading trajectories shown in Fig.3(a). The attachment rate is then calculated by linear fitting of the ensemble average of the nucleus size change according to eq. (4). $f^+_{MD}$ for all the Sm compositions studied in this paper is given in Fig. 3(b). With 5 at.% Sm doping, the Al attachment rate decreases by one order of magnitude.

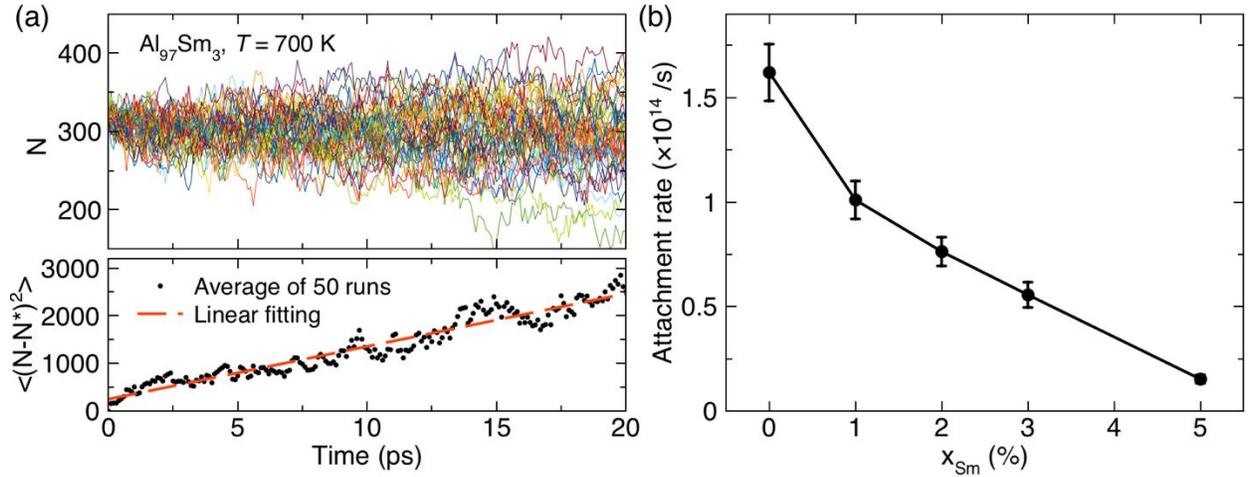

Fig. 3 The measurement of attachment rate at the critical size. (a) The nucleus size versus time for the isoconfigurational ensemble with 50 MD runs. Each color indicates a MD trajectory. (b) the ensemble average of $|N(t) - N^*|^2$. The dashed line indicates the linear fitting to derive the attachment rate. (b) The attachment rates obtained by the MD measurement as a function of Sm doping concentration. The error bar is obtained by repeating the isoconfigurational MD for different critical nucleus collected in PEM.

With all the quantities available as listed in Table 1, we computed the nucleation rate using eq. (3) for fcc-Al in Al$_{1-x}$Sm$_x$ undercooled liquid at 700 K. With 1% Sm doping, the fcc nucleation rate does not change much compared to the one in pure Al. However, when doping concentration increases up to 5%, the nucleation rate drops dramatically by almost 25 orders of magnitudes. Such significant effect is mostly associated with the increase in the free energy barrier with minor contribution from suppression of the attachment to the fcc nucleus caused by Sm doping. This dominant mechanism of the increased free energy barrier caused by the minor Sm doping in the moderately undercooled liquid is different in the glass state studied in Ref. [20] where the kinetics barrier dominates the nucleation process. The results from current study, together with the data measured in Ref. [20] are both critical to describe the nucleation rate from high temperature regime to the glass state, which is shown in the Supplementary Material.

Table 1 The quantities contributing to the fcc nucleation rate at T=700 K as a function of the Sm doping concentration ($x_{Sm}$): the chemical potential difference between fcc-Al and Al in AlSm undercooled liquid ($\Delta\mu$); the critical nucleus size (N$^*$); the free energy barrier $\Delta G^*$; the attachment rate ($f^+$) and the nucleation rate (J).

| $x_{Sm}$ | $\Delta\mu$ (eV/atom) | $N^*$ | $\Delta G^*$ ($k_B T$) | $f^+$ ($s^{-1}$) | $J$ ($m^{-3} s^{-1}$) |
|---|---|---|---|---|---|
| Al | -0.0235 | 155 | 30.2 | $1.6 \times 10^{14}$ | $7.6 \times 10^{27}$ |
| Al$_{99}$Sm$_1$ | -0.0229 | 165 | 31.4 | $1.0 \times 10^{14}$ | $1.3 \times 10^{27}$ |
| Al$_{98}$Sm$_2$ | -0.0220 | 227 | 41.3 | $7.6 \times 10^{13}$ | $4.1 \times 10^{22}$ |
| Al$_{97}$Sm$_3$ | -0.0212 | 310 | 54.6 | $5.6 \times 10^{13}$ | $4.5 \times 10^{16}$ |
| Al$_{95}$Sm$_5$ | -0.0191 | 546 | 86.3 | $1.5 \times 10^{13}$ | $1.5 \times 10^2$ |

With PEM, we also investigate the effect of the Sm dopant on the sub-critical stage of the Al nucleation. To show the effect of the trapped Sm on the embryo, we perform a comparative PEM studies with pure-Al and Sm-doped embryo in Al$_{97}$Sm$_3$ system at *T*=700K. In Fig. 4 (a), we first perform PEM simulation with a pure Al embryo. 10 independent MD simulation are performed and all of samples nucleate within 30 *ns*. In Fig. 4 (b) and (c), we replace one Al atom with Sm atom in the embryo and perform PEM simulation again. It can be seen that no nucleation happens within 50 *ns*. Most importantly, once the spring of the Sm is removed at 50 *ns* so that the constrained Sm is free to diffuse, the nucleation starts to happen in all the samples. It demonstrates that at *T*=700K any trapped Sm atom on the sub-critical nucleus will significantly increase the free energy barrier of forming the critical nucleus. When such free energy barrier is the dominant factor of the nucleation and much larger than the Sm diffusion barrier, the nucleation can only occur when the trapped Sm diffuses away. Therefore, the system will only form a pure-Al critical nucleus. This scenario is also consistent with the fact that all the as-formed critical nucleus in the PEM simulations contains no Sm.

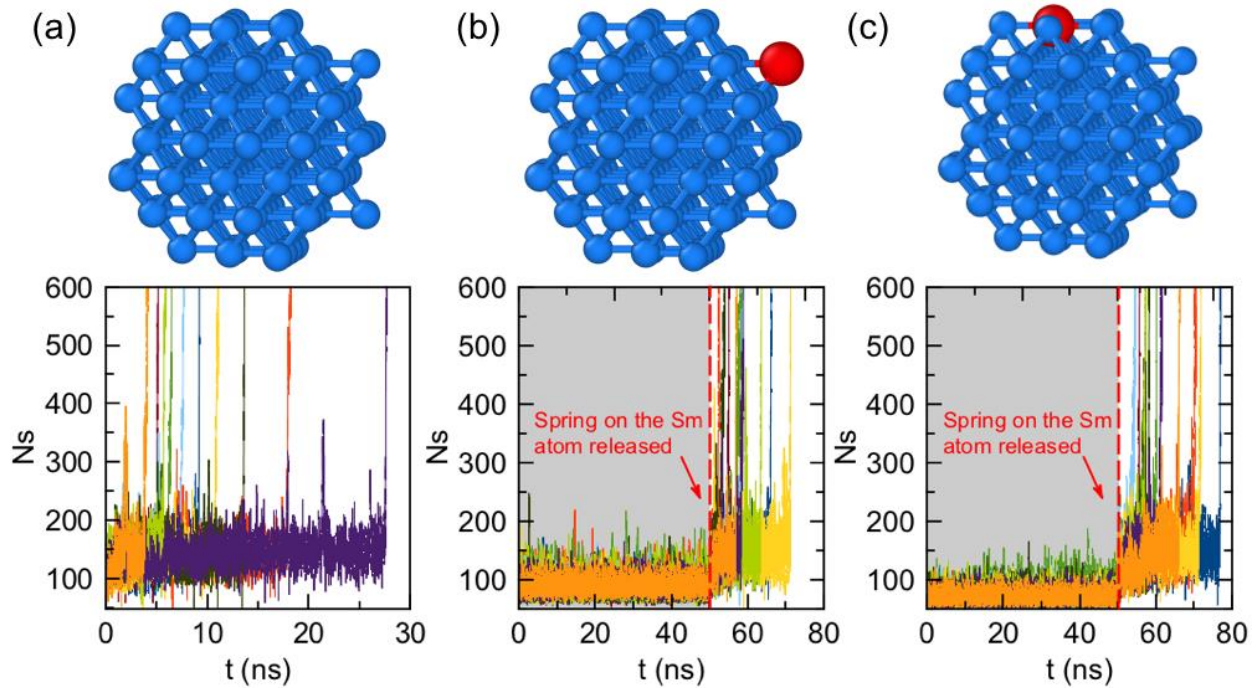

Fig. 4 PEM simulation with the Sm-doped embryos. (a) PEM simulations with pure Al embryo in the $Al_{97}Sm_3$ undercooled liquid. The upper panel shows the atomic configuration of the spring constrained embryo. Blue represents Al atoms, while red is Sm atom. The lower panel shows the nucleus size (Ns) as a function of the simulation time. 10 independent simulations are performed for the embryo. The rapid increase of Ns indicates the occurrence of the nucleation. All of samples nucleate within 30 *ns*. (b) and (c) shows PEM simulation with two different embryo configurations by replacing one Al with Sm. No nucleation can be observed within the first 50 *ns*. After the spring on the Sm is released at 50 *ns*, nucleation starts to happen.

In summary, the homogeneous fcc nucleation was studied in MD simulation of moderately undercooled $Al_{1-x}Sm_x$ (x=0%, 1%, 2%, 3%, 5%) liquids using the persistent embryo method. Using the MD simulation data, the nucleation rate was quantitatively computed in the framework of CNT. The Sm doping is found to significantly increase the fcc nucleation barrier, decrease the driving force of the fcc nucleation and suppress the Al attachment rate, which together lead to a huge decrease of the nucleation rate up to 25 orders of magnitudes with only 5% Sm dopant at same temperature. Quantifying the effect of the Sm doping on the Al nucleation can be useful for the understanding of the solidification pathways and further development of the phase-field modeling for the phase selection process.

**Acknowledgements**

Work at Ames Laboratory was supported by the US Department of Energy, Basic Energy Sciences, Materials Science and Engineering Division, under Contract No. DEAC02-07CH11358, including a grant of computer time at the National Energy Research Supercomputing Center (NERSC) in Berkeley, CA. K.M.H. acknowledges support from USTC Qian-Ren B (1000-Talents Program B) fund. The Laboratory Directed Research and Development (LDRD) program of Ames Laboratory supported the use of GPU computing.